\documentclass[12pt,preprint]{aastex}
\slugcomment{Accepted for AJ (Feb. 07 issue)}
\shorttitle{Parallaxes of Planetary Nebulae}
\shortauthors{Harris et al.}

\begin{document}

\title{Trigonometric Parallaxes of Central Stars of Planetary Nebulae}

\author{Hugh C. Harris\footnote{
Guest observer, Kitt Peak National Observatory, National Optical
Astronomy Observatories, operated by the Association of Universities
for Research in Astronomy, Inc., under cooperative agreement with
the National Science Foundation.},
Conard C. Dahn, Blaise Canzian, Harry H. Guetter, S.K. Leggett\footnote{
Current address Gemini Observatory, 
670 North A`ohoku Place, Hilo, HI 96720-2700},
Stephen E. Levine, Christian B. Luginbuhl, Alice K.B. Monet, David G. Monet,
Jeffrey R. Pier$^1$, Ronald C. Stone, Trudy Tilleman, Frederick J. Vrba,
Richard L. Walker}

\affil{U.S. Naval Observatory, 10391 W. Naval Observatory Rd,
Flagstaff, AZ 86001;  hch@nofs.navy.mil}

\begin{abstract}
Trigonometric parallaxes of 16 nearby planetary nebulae are presented,
including reduced errors for seven objects with previous initial results
and results for six new objects.  The median error in the parallax is
0.42 mas, and twelve nebulae have parallax errors less than 20 percent.
The parallax for PHL932 is found here to be smaller than was measured
by Hipparcos, and this peculiar object is discussed.
Comparisons are made with other distance estimates.
The distances determined from these parallaxes
tend to be intermediate between some short distance estimates and other
long estimates;  they are somewhat smaller than estimated from spectra
of the central stars.  Proper motions and tangential velocities are
presented.  No astrometric perturbations from unresolved close companions
are detected.
\end{abstract}

\keywords{Astrometry --- ISM: Planetary Nebulae: General ---
Stars: AGB and Post-AGB --- Stars: Distances}

\section{INTRODUCTION}

Distances of planetary nebulae continue to be quite uncertain despite
the increased quality and quantity of data available in recent years.
In part, the variety and complexity of nebular structures seen in
high resolution images suggest that some methods of determining
distances incorporate simplifying assumptions that are not realistic.
These methods may be affected by errors even when detailed models
are made of each nebula.

Measuring trigonometric parallaxes of the central
stars of planetary nebulae (CSPN) can, in principle, provide a check on
other methods for at least the nearest nebulae.  Until recently,
however, planetary nebulae have eluded successful parallax measurements.
The Yale Parallax Catalogue \citep{van95} lists 24 CSPN with measured
parallaxes, but only one (the central star of NGC~7293) has an error
less than 8~mas\footnote{
The parallax of the central star of NGC~7293 was marginally detected in
several studies.  The best result, according to the Yale Catalogue,
was $5 \pm 5$ mas \citep{dah73, har80}.  Below we obtain 4.6 $\pm$ 0.5
mas, demonstrating the improved precision that can now be achieved.
},
so these measurements are not significant.
The Hipparcos satellite targeted 19 CSPN.  Unfortunately,
the faintness of the stars -- near Hipparcos' magnitude limit --
caused the results to have errors larger than the $\pm$1~mas typical for
most Hipparcos parallaxes.  When combined with the small parallaxes
that most of these CSPN have, the results were disappointing;
only four stars had an error less than 50\% of the parallax
\citep{ack98, pot98}.  (Below we remeasure one, and find a
parallax smaller than the Hipparcos value by $2\sigma$.)
Therefore, Hipparcos results have not been relied
upon for establishing a planetary nebula distance scale.

Using the advantages offered by cameras with CCD detectors (high
signal/noise, high dynamic range, high quantum efficiency), the U.S.
Naval Observatory began observing some CSPN for parallaxes in 1987.
Initial results were reported at IAU Symposia 155 \citep{pie93}
and 180 \citep{har97}.  Continued observations now give greatly
improved and expanded results that are the subject of this paper.

A parallax for one CSPN has been measured using the fine guidance
sensors on Hubble Space Telescope \citep{ben03},
and data are currently being obtained for three other CSPN.
Distances using the expansion method are also being measured with HST and
the VLA by Hajian (2006) and collaborators.  One other ground-based
parallax study of three PNe (Guti\'errez-Moreno et al. 1999) gave results
 with less precision than is desirable.  A few CSPN have companions for
which a spectroscopic distance estimate has been made \citep{cia99}.
Finally, spectroscopic analysis of the CSPN has given distance estimates
(through determination of the gravity of the CSPN) for many nebulae
\citep{nap99, nap01}, some of which are in common with this paper.
In Section 2 we present our astrometry, photometry, and parallax results,
and in Section 3 we compare our results with those of other methods.

\section{OBSERVATIONS}

\subsection{Astrometry}

The USNO parallax program began using CCD detectors in 1985.
The observing and data processing techniques are described by Monet
et al. (1992) and Dahn et al. (2002).  Originally, the detector used
was a Texas Instruments 800$\times$800 (TI800) CCD, and observations of
nine PNe were begun in 1987 using this camera.  In 1992, a larger format
Tektronix 2048$\times$2048 (Tek2048) CCD was put into use, and, in 1995,
use of the smaller TI800 camera was discontinued.
The much wider field of the Tek2048 CCD (11 arcminutes, compared
with 3 arcminutes for the TI800 CCD) and the greater full well
of the Tek2048 pixels (giving good signal-to-noise data over a
greater dynamic range of bright and faint stars) are both
advantageous for parallax work;  they allow selection of better
reference stars (more stars, more distant stars, and stars with
a more symmetrical distribution around the parallax star)
which is important in sparse fields.
All observations have been made with a wide-$R$ filter.

The precision of USNO parallax results has improved steadily
during the 20 years of using CCD detectors \citep{har05},
owing to a combination of improved data quality control
and the accumulation of large numbers of CCD frames over many
nights and observing epochs.
The precision of parallaxes is now better than 0.5~mas routinely,
and is better than typical parallaxes from Hipparcos by factors
of 2--3.  For stars like CSPN at distances of hundreds of parsecs,
this improvement in precision is a crucial factor to obtain
significant results for many objects.  In fact, as a class,
planetary nebulae are the most challenging objects observed at
USNO for obtaining scientifically useful results.

The results of our astrometric data analysis are shown in Table 1.
Parallaxes determined with the TI800 CCD have a median error
in the relative parallax of 0.56 mas, while parallaxes determined
with the Tek2048 CCD have a median error of 0.42 mas.
(For three stars, observations are continuing, so the latter
error is expected to drop further as more data are obtained.)
Three PNe (A24, A29, and A31) were begun with the TI800 camera,
but satisfactory results were not obtained before use of that
camera was discontinued.  All three have small parallaxes
where high precision is essential for useful results.  Two of
these three have now been completed with the Tek2048 camera,
while the third (A29) has been dropped from the program.
Three objects have results from both cameras, and one has a
further (but not yet final) parallax observed with HST \citep{ben03}.
These three provide a test for consistency, and provide one of the
few available assessments for accuracy:  for Sh2-216, PuWe1, and
NGC~6853 results from the two cameras agree within $0.4\sigma$,
$0.8\sigma$, and $1.9\sigma$, respectively; for NGC~6853, the result
from the TI800 and Tek2048 cameras differ from HST by $0.8\sigma$
and $2.5\sigma$, respectively.  The residuals for NGC~6853 also show
a correlation with seeing that indicates a problem with the images
of the central star, such as might be caused by the known faint
companion star (see notes below).  While the problem with NGC~6853
is not presently understood for certain, overall this level of
agreement between different cameras is good.

A correction for the finite distances of the reference stars
must be applied to convert each observed relative parallax
to absolute parallax.  This correction is important for PNe
with distances of several hundred pc, and becomes increasingly
important as the error in the relative parallax drops.
This correction is shown as $\Delta \pi$ in Table 1.
It has been determined both from spectroscopy and photometry of the
reference stars.  The use of $BVI$ photometry for determining the
corrections is done for all USNO parallax fields, and
is described by Dahn et al. (2006, in preparation).
In addition, for the PNe in this paper, spectra of many reference
stars and spectral classification standard stars
were taken at Kitt Peak National Observatory (4-m telescope,
R/C spectrograph, CCD camera, 7\AA\ resolution) and at USNO (1-m
telescope, fiber-fed bench spectrograph, CCD camera, 8\AA\ resolution).
From these spectra, the strengths of 17 spectral features (H lines,
Ca K, G and CN bands, MgH, MgI, Na D, CaI, and several Fe features)
were measured and were used to determine the spectral type and
luminosity class of each star and to check for any peculiarities
(e.g. binary star, metal-poor star, emission lines, etc.).
Particularly in fields at low galactic latitude, where reddening
can be large and very different from star to star, the combination
of spectral indices with photometric data is helpful for determining
a reliable correction to absolute parallax.

The combined data give an estimate of the reddening and
distance of each reference star, and allow the rejection from the
astrometric solutions of any nearby (or uncertain) reference stars.
Finally, 4 to 8 reference stars have been used in the final solutions
for the PNe observed with the TI800 camera, and from 4 to 32 stars
for those observed with the Tek2048 camera.
Note that for fields close to the galactic plane (particularly Sh2-216,
and to a lesser extent NGC~6853), reddening is large enough to make the
distance of each reference star more uncertain, adding to the error
in $\Delta \pi$.  For all fields, Table 1 shows that the reference stars
are typically at distances of 1--2 kpc.  At these distances,
$\Delta \pi$ is fairly small and reasonably well-determined.  The median
error in $\Delta \pi$ (the important quantity for this correction) is
0.17~mas, whereas the median error in $\pi_{rel}$ is 0.42~mas.
Therefore, for most of these PNe, the error in $\Delta \pi$ contributes
only a small amount to the final error in the absolute parallax.
Note that the use of {\it faint} reference stars is an important factor
in keeping this contribution to the total error small.  With {\it faint}
reference stars, the error in the observed relative parallax usually
dominates the total error.

\subsection{Photometry}

Photometry with $B$, $V$, and $I$ filters has been obtained
for the CSPN in this paper and their reference stars
with the USNO 1-m telescope during 1994-2005.
Standard stars \citep{lan92} were observed each night.
The filters have passbands close to the standard Johnson
(for $B$ and $V$) and Cousins (for $I$) passbands, and small
color terms were determined and applied each night for these
very blue program stars.  The error of a single observation
is typically 0.03 in $V$ and 0.03 in $B-V$ and $V-I$.

Results are shown in Table 2 for the central stars.
For A74, Table 2 shows the star to be notably brighter and bluer
than found by Ciardullo et al. (1999) using HST.  For the remaining
8 stars in common, we find a small mean difference of 0.066 $\pm$ 0.014
brighter in $V$ and 0.036 $\pm$ 0.016 redder in $V-I$ than found
by Ciardullo et al.
Nebular contamination should not be causing significant errors
for our CCD photometry of these CSPN with large nebulae, nor for
the HST photometry.  More likely these differences may be related
to transforming magnitudes from the HST filter system used by
Ciardullo et al. to the Johnson/Cousins system used here.

\subsection{ Notes for individual central stars}

NGC~6720 (Ring Nebula) -- No additional data; result repeated
from IAU Symp. 180.

NGC~6853 (Dumbbell Nebula)  -- Distances to reference stars are
somewhat uncertain because of uncertain reddening along the line of
sight beyond the nebula.
Our results with two cameras differ from each other by $2\sigma$.
We see systematic residuals vs. seeing that might be caused by a
close pair of stars.  A star with $V \sim 18.7$ and 1$\arcsec$
separation is reported \citep{cia99}, and may be the cause of
our correlation with seeing.  As a result, the ground-based
astrometry may be adversely affected by the nearby star.
Therefore, we adopt a weighted mean of the two USNO determinations
given in Table 1, but with an error estimate inflated to account for
the possible interference:  $\pi{\rm(USNO)} = 3.17 \pm 0.47$~mas. 
(This error inflation is based on subjective review of various
parallax solutions using different subsamples of data.)
An initial result using Hubble Space Telescope \citep{ben03} is
$\pi{\rm(HST)} = 2.10 \pm 0.48$~mas, and one additional HST epoch
is being obtained.  For an overall best current parallax, we adopt
the weighted mean of $\pi{\rm(USNO)}$ and $\pi{\rm(HST)}$ of
of $2.64 \pm 0.33$~mas.

NGC~7293 (Helix Nebula) -- Result improved from IAU Symp. 180.

A7 -- Result improved from IAU Symp. 180.  A faint, red companion star
at 0.9$\arcsec$ separation is possibly a physical companion
\citep{cia99}.

A21 -- No additional data; result repeated from IAU Symp. 180.

A24 -- Result improved from IAU Symp. 180 with Tek2048 camera,
and the earlier preliminary measurement is omitted here.
A red companion star at 3.4$\arcsec$ separation is not a physical
companion \citep{cia99}.

A29 -- Result from IAU Symp. 180, 2.18 $\pm$ 1.30 mas, noted
``Insufficient data'' to obtain a reliable parallax,
and star dropped from USNO program.
Result from Guti\'errez-Moreno et al. (1999), 3.3 $\pm$ 1.2 mas.
Weighted mean of these two measurements gives 2.8 $\pm$ 0.9 mas,
but these results are not included in the remainder of this
paper (see Sec. 3).

A31 -- Result improved from IAU Symp. 180 with Tek2048 camera,
and the earlier preliminary measurement is omitted here.
A red companion star at 0.26$\arcsec$ separation is likely to be
a physical companion and, if so, implies a distance $<440$~pc
\citep{cia99};  the parallax in Table 1 is 1.5$\sigma$ smaller
than implied by such a close upper limit to the distance.

A74 -- No additional data; result repeated from IAU Symp. 180.
A star at $3.5 \arcsec$ separation with $V=18.44$, $V-I=1.04$
is too distant to be a physical companion.
Photometric difference between this paper and Ciardullo et al.
(1999) suggests real variability.

Sh2-216 -- Result improved from IAU Symp. 180; results with two
cameras agree well.  Distances to reference stars are uncertain
because of the large reddening along the line of sight beyond
the nebula; however, the parallax of Sh2-216 is sufficiently large
that this uncertainty is a small fraction of the total parallax.

PuWe1 -- Result improved from IAU Symp. 180; results with two
cameras agree well.
A red companion star at $5 \arcsec$ separation is itself a resolved
pair, but probably not physical companions at the same distance as
PuWe1 \citep{cia99}.

DeHt5 -- New result; observations continuing.  We see systematic
residuals vs. seeing that might be caused by a close pair of stars.
Frew \& Parker (2006) find nebula may be ionized ISM.

HDW4 -- New result.  Frew \& Parker (2006) find nebula may be
ionized ISM.

Ton320 -- New result.

RE1738+665 -- New result.  Frew \& Parker (2006) find nebula
may be ionized ISM.

PHL932 -- New USNO result; preliminary result due to data over
a small epoch range (Table 1); observations continuing.
The parallax from Hipparcos is $9.12 \pm 2.79$~mas, but the USNO
parallax in Table 1 is smaller by $2\sigma$.  A weighted mean of
the Hipparcos and USNO results is $3.63 \pm 0.61$~mas; we use
only the USNO results hereafter, as discussed in Sec. 3\footnote{
The central star of PHL932 appears to have a firm $3\sigma$
parallax observed by Hipparcos \citep{ack98}, but is found here
to have a parallax smaller by a factor of 2.7 ($2.0\sigma$).
We have run numerous solutions of our data, varying the reference
stars and the frames included in the solutions, and these experiments
appear to rule out a parallax anywhere near the large Hipparcos value.
Is the Hipparcos parallax just a $2\sigma$ event, or an outlier
caused by a systematic or non-Gaussian error to be rejected?
Emission from the nebula of PHL932 should not
affect the the Hipparcos measurement \citep{ack98}.
The Hipparcos error is typical for this star near the faint
limit for Hipparcos:  the Hipparcos and Tycho Catalogues (Vol. 1,
Table 3.2.4) give a median $\sigma_{\pi} = 2.98$~mas for $Hp = 11-12$
and $\sigma_{\pi} = 4.35$~mas for $Hp > 12$ at $\beta < 10\deg$,
so for PHL932 ($Hp = 12.03$), we expect $\sigma_{\pi} = 3.5$~mas.
Only 0.5\% of stars in the Hipparcos catalog are as faint as PHL932;
the statistical properties of these faint stars and the accuracy of
their parallaxes and error estimates are not as well known as the bulk
of the brighter stars in the catalog.
Of the 19 CSPN observed by Hipparcos, only two are fainter than
PHL932, and they both have larger Hipparcos errors (as do some of
the brighter CSPN).  Five of the 19 have negative Hipparcos parallaxes,
and one might use this distribution to try to access the accuracy of
the parallax and error values, but doing so would be difficult without
knowing more about their true parallaxes.  A remote possibility is
that an astrometric perturbation has affected the Hipparcos
measurements, and perhaps the USNO measurements too in an opposite
sense by chance;  if so, it should be identifiable with an additional
couple of years of data.  At present, we do not understand what has
caused the Hipparcos and USNO parallaxes to differ.
}.
Several peculiarities, see discussion in Sec. 3.

PG1034+001 (WD1034+110) -- New result; preliminary result due to
data over a small epoch range (Table 1); observations continuing.
Large nebula with a larger outer halo \citep{hew03, rau04}.
Frew \& Parker (2006) find nebula may be ionized ISM.

\section{DISCUSSION}

Resulting combined parallaxes from Table 1 (and one from the
literature) and distances are shown in Table 3.  In constructing
Table 3, several other parallax determinations from the literature
have been ignored.  The goal of our program has been to reach parallax
errors much smaller than 1~mas, both to determine accurate distances
and absolute magnitudes and to minimize possible Lutz-Kelker-type
biases and bias corrections discussed below.  With this goal in mind,
we have chosen to ignore all parallaxes with estimated errors
greater than 1.0~mas\footnote{
The omitted measurements include that for PHL932 from Hipparcos,
9.12 $\pm$ 2.79 mas (see discussion in Sec. 2.3),
for A21 from Guti\'errez-Moreno et al. (1999), 1.9 $\pm$ 1.3 mas,
and for NGC~7293 from Dahn et al. (1973), 5 $\pm$ 5 mas.
Also omitted are the three measurements from our own CCD program
with the TI800 camera from Harris et al. (1997) that were noted as
``Insufficient data, preliminary result'' (based on the number of
frames, the distribution of parallax factors, the quality of the
reference stars, and the stability of the parallax solutions)
for A24, 3.11 $\pm$ 1.12 mas, for A29, 2.18 $\pm$ 1.30 mas, and
for A31, 4.75 $\pm$ 1.61 mas.  Subsequently A24 and A31 have been
measured with the Tek2048 camera and appear in Tables 1 and 3.
}
in an effort to keep the accuracy of the measurements close to
the precision errors given in Table 3.  Omitting these results is
an arbitrary procedure, and undoubtedly omits some valid results,
but including them with appropriate weights would not change the
Table 3 averages much.

The precision of the parallax determinations for most objects in
Table 3 is improved considerably from earlier results \citep{har97}:
two stars have a result that is greatly improved, five stars are
somewhat improved, three stars have no additional data so are
unchanged, one star has been dropped, plus an additional six stars
have been added.  Of the five stars from Harris et al. (1997)
that had sufficient data to have a believable result in 1997
but have improved results now, one star (A7) has a parallax larger
by $2\sigma$, one star (PuWe1) has a parallax larger by $1\sigma$,
and the other three have parallaxes changed only slightly.
In 1997, four stars had a parallax significant at $5\sigma$ or greater,
now 12 stars have a parallax with this level of significance.

The last column in Table 3 gives the tangential velocity determined
from the relative proper motion in Table 1 and the distance,
uncorrected for solar motion.  The median tangential velocity is
29 km~s$^{-1}$, the extreme is 88 km~s$^{-1}$, and only one star
(PG1034+001) has a tangential velocity above 55 km~s$^{-1}$.
These values by themselves are consistent with all stars being
thin disk stars.  However, using the new distance in Table 3,
PG1034+001 has a tangential velocity sufficiently high that it
could be a member of the thin disk or the thick disk (see note below).

Table 3 includes our determination of E($B-V$) for the central star
from our observed $BVI_{\rm C}$, assuming ($B-V$)$_{\rm 0} = -0.38$ and
($V-I$)$_{\rm 0} = -0.45$, and the derived M$_V$.  The results
are consistent with most stars having $6.0<{\rm M}_V<7.5$.
Notable exceptions are the central stars of A24, HDW4, and RE1738+665,
which appear to be fainter,
and the central star of PHL932, which is much brighter (see below).
Phillips (2005a) suggested that in evolved PNe with low radio surface
brightness (log T$_{\rm B}$ in the range -0.5 to -4.0), most central
stars have $<M_V>=7.05$ with a range $\pm 0.5$.  There are 11 stars
in Table 3 within the range of T$_{\rm B}$, and they have 
$<{\rm M}_V>=7.07$, in excellent agreement with Phillips.
However, the range is 4.43 (PHL932) to 9.43 (HDW4).
Of the 11 stars, four differ from 7.05 by 1 mag or more, so are well
outside the range in which Phillips suggests that most stars lie.
These outliers indicate the real range of $M_V$ is closer to $\pm 2$
than $\pm 0.5$ mag.  Therefore, assuming a constant $M_V$ to determine
distances would result in significant errors, at least for these 11 stars.
However, considering the peculiarity of PHL932 (discussed below) and the
possibility that HDW4 and RE1738+665 are not true PNe (noted in Sec. 2.3),
most of the remainder of our sample do satisfy Phillips' suggestions of
nearly constant $M_V$ with a small true dispersion.  We may be able to
determine the true dispersion in $M_V$ from these data after the nature
of the nebulae for these outliers is clarifed.

In Table 4, a comparison is made between the distances determined here
and those from several other discussions of PNe distances.  (The
comparison papers are selected as particular methods or as representative
review papers, and are not intended to give a thorough discussion of
the many distance scale determinations in the literature.)
In general, we find larger distances than have been determined in some
older studies (although PHL932 is an exception to this trend).
For example, we have observed eight of the 11 stars listed by Terzian
(1993, his Table 3) as having distances $<300$ pc.  Terzian's summary
of nearby planetary nebulae was writen prior to the availability
of any useful parallax data.  We find only two of these eight stars
to be within 300 pc, while we find DeHt5 to be at a distance of 300 pc
compared to Terzian's 400 pc.  The comparison is shown in Figure 1.
Overall, the nearby PNe are not as nearby as believed by Terzian
(1993).  A comparison with Pottasch (1996) is shown in Figure 2,
and this agreement is better.  However, he included the preliminary
parallax results for seven stars available then \citep{har97},
so some improved agreement is expected.

Table 4 includes comparison with the statistical distances found
by Cahn et al. (1992) (known to give ``short'' distances) and by
van de Steene \& Zijlstra (1994) and Zhang (1995) (both know to
give ``long'' distances).  Inspection of Table 4 shows that
this paper gives distances that agree better with the short scale
for some objects and better with the long scale for others;
overall this paper gives an intermediate distance scale.
However, the scatter in these comparisons with all three sources
is large.  With the large scatter, the small sample of objects in
common makes quantitative comparisons uncertain.  Furthermore,
the poor performance of these statistical distances
for evolved, low surface-brightness nebulae has been noted before
(e.g. Ciardullo et al. 1999; Phillips 2005a; 2005b), as has the
dichotomy between the short and long distance scales.  Unfortunately,
parallax measurements are necessarily confined to nearby distances
where only evolved nebulae have been found, so our sample includes
none of the young, high surface-brightness nebulae for which the
statistical distance procedures probably give more reliable results.
Therefore, a quantitative comparison between the parallax and
statistical distances may be misleading when applied only to these
evolved nebulae.

A comparison between spectroscopic distances (from analysis of the
CSPN with non-LTE model atmospheres) and distances determined by
trigonometric parallax was done by Napiwotzki (2001).  He showed that
spectroscopic distances were greater than trigonometric distances by
a factor of 1.55, and he argued that a bias in the trigonometric
distances from a combination of sample selection and a Lutz \& Kelker
(1973) bias accounted for the discrepancy.  Therefore, he concluded,
the parallaxes corrected for his estimate of the bias supported the
longer spectroscopic distance scale.  Is this comparison still valid?
First, Napiwotzki included the parallax of the central star of PHL932
from Hipparcos, and this paper finds the parallax to be smaller
by a factor of 2.7.  The remaining seven stars used by Napiwotzki, using
parallaxes from Harris et al. (1997), have spectroscopic distances
greater by a factor of 1.4.  The new parallax values given in Table 3
have not changed much, but the errors are reduced, as noted above.
We can now add four stars with new parallaxes and previous spectroscopic
distances \citep{nap01, nap99, pot96, wer95}.  The comparison is shown
in Figure 3.  Three stars have a spectroscopic distance smaller than
the distance determined by parallax, one star has good agreement,
and 11 stars have a larger spectroscopic distance.  The weighted mean
distance ratio, using the error estimates for both distances, has
spectroscopic distances larger than trigonometric distances by a factor
of 1.3.

Napiwotzki (2001) used Monte Carlo models to estimate the bias in the
parallax-based distances.  A correction for the bias is significant when
the parallax errors are an appreciable fraction of the parallax values,
but it becomes small or insignificant when the parallax errors are
reduced, as Napiwotzki's Figure 2 illustrates.  Several of the stars had
small parallax errors then, and a bias correction would not have
accounted for the discrepancy with spectroscopic distances.
Now, with parallax errors reduced, a bias correction is also reduced.
As Figure 3 shows, five central stars (NGC~7293, A31,
Sh2-216, PuWe1, and DeHt5) have spectroscopic distances larger than
the distances determined by parallax by more than $1\sigma$ in both
errors {\it and} have small parallax errors;  the mean $\sigma/\pi$
is 0.12 for these five stars.  The parallax data suggest these stars
may have their spectroscopic distances overestimated.

An estimate for bias in the distances and absolute magnitudes
derived from parallax measurements can best be made using Monte
Carlo models like those of Napiwotzki (2001) or Smith (2003), for
example.  In many applications, models like these can quantify the
effects of the Lutz-Kelker bias as well as additional effects from
sample selection, and they can include a non-uniform space distribution
of the observed objects.  However, the sample of CSPN in this paper
is not easily modeled because of the poorly defined criteria for
selecting objects for our observing program.
We have tended to include objects believed (from literature estimates)
to be nearby, but the unreliability of these prior distance estimates
has resulted in a few objects like A7 and A74 turning out to be at much
greater distances than originally expected.
Because all our parallax results have been published, there is no
selection based on minimum parallax or maximum fractional error
such as often occurs when extracting a sample from a large catalog
of parallax results.  A faint magnitude limit is not significant,
because we routinely include fainter stars in our program, so
magnitude and reddening can be omitted from the model.  We have
rejected a few bright CSPN (where the CSPN has a brighter companion)
from our program thus far, but this selection will not introduce
significant bias.

As an indicator of the approximate bias in the results in Table 3,
Figure 4 shows two models of a disk population of objects with
a scale height of 250~pc and with parallax measurements
with rms errors of 0.42~mas to match the errors in Table 3.
In the models, 200,000 objects are placed with random positions,
including the exponential scale height in z, in a box 3 kpc on a side
with the sun at the center.  (Fewer objects are plotted in Fig. 4
for clarity in the plot.)  The model in the top panel has parallax
objects selected randomly.  The reduced density of objects in the
upper right corner is due to incompleteness beyond 1.5 kpc caused
by the box size.  The top panel shows that the bias (a Lutz-Kelker-type
bias, but modified by the non-uniform density of the disk distribution)
causes the parallax distances to be underestimated by 19\% (and the
absolute magnitudes to be 0.38 mag too faint) for objects with
estimated parallax errors of 20\% of the measured parallax.
In contrast, the model in the lower panel has an added selection of
likely-nearby objects, attempting to match our sample selection.
Objects were included in this model sample if their estimated distance
was $<550$~pc after adding a 30\% rms error to the true model distance.
(That is, a gaussian error was added to their true model distance to
give a pseudo-distance, to mimic the rough estimated distances available
from the literature for selecting objects to include in our program.
Then, if this pseudo-distance was $<550$~pc, the object was included
in the model sample.  These parameters were chosen to give a model
sample with a median measured parallax of 2.54~mas and a median relative
parallax error of 17\%, values that closely match the actual parallax
sample in Table 3.)
The lower panel shows that the bias causes the parallax distances to be
underestimated by 5\% (and the absolute magnitudes to be 0.11 mag
too faint) for objects with relative parallax errors of 12\%,
and the bias then drops to about zero for objects with relative parallax
errors of 20\%.  With 12 objects in Table 3 having a parallax error
less than 20\% of the parallax, Fig. 4 suggests that the mean bias
in their distances is probably about 5\% or less.

The distance of PG1034+001 of $211^{+26}_{-22}$~pc is greater than the
previous spectroscopic determination of $155 \pm 50$~pc \citep{wer95}.
This has two ramifications for the discussion about the size
of the PN and the galactic orbit of the central star \citep{rau04}.
First, the inner nebula with a diameter of 2 degrees has a linear
diameter of 7 pc, the fainter halo with a diameter of 6x9 degrees has
a diameter of 21x32  pc, and the outermost fragmentary shell has a
diameter even larger.  This nebula would be the largest known if
it is a true PN.  Its large size and a difference in radial velocity
between the nebula and the central star has led to debate about the
origin of the nebulosity \citep{chu04, fre06}.  However, some
interaction with surrounding interstellar material is expected for large,
old, evolved PNe \citep{twe96} such as are discussed in this paper.
Second, the proper motion in Table 1 is in good agreement with previous
proper motion measurements \citep{rau04}.
With the revised distance, the tangential velocity is now increased
to 88 km~s$^{-1}$, the galactic orbit calculated from the proper motion
and radial velocity will be altered, and the conclusion drawn by
Rauch et al. about its origin in the thin disk may be changed.

The distance of RE1738+665 of $169^{+13}_{-11}$~pc is slightly closer
than the $200 \pm 35$~pc derived spectroscopically \cite{bar94}.
This distance supports the gravity (log $g$ = 7.8) and mass
($\sim 0.62 M_{\odot}$) of the central star \cite{bar03}
being somewhat higher than most CSPN.  A relatively high mass
was proposed \citep{twe95} to account for its X-ray emission
discovered in the Rosat survey.  The emission was noted as unusual
for such a hot, hydrogen-rich star, and is still unusual, despite
the fact that the temperature derived now by Barstow et al. for
RE1738+665 (and for other DA stars) is less extreme than considered
by Tweedy.

The nature of PHL932 is problematic in several respects.  First, we
find a smaller parallax than measured by Hipparcos (see Sec. 2.3).
The larger distance given in Table 3 helps
resolve some of the discrepancy in log~$g$ for the central star
between the observed gravity \citep{men88} and the gravity
deduced from the temperature and an assumed (normal) mass
\citep{pot98}.  Nevertheless, the central star is a sdB star
on the extended horizontal branch \citep{lis05}, requiring extreme
evolution to produce a PN.  DeMarco et al. (2004) find the radial
velocity is variable.  One possibility is that the nebula is
ionized ISM \citep{fre06}, not a true PN.  Alternatively,
the true central star may be fainter and unresolved and not yet
observed -- with M$_V = 4.4$ for the sdB star in Table 3,
a CSPN with a typical M$_V \sim 7$ could easily be hidden
and only detectable at UV wavelengths.  The radial velocity
variations could be caused by the real CSPN, or there could be
a more complicated triple system.

In observing nearby stars with accurate astrometry, it is sometimes
possible to detect perturbations from an unresolved close
companion with a period of the length of the observations
or less, if such a companion exists.  The USNO parallax program
has discovered a number of such companions around nearby
red dwarfs and white dwarfs.  No perturbations are apparent
for the 16 CSPN in this paper.  These data rule out companions
with periods less than 5--10 yr and with masses large enough to
produce an apparent reflex motion of the central star of $>1$~mas.
In practice, early-M dwarf companions with these periods would be
detected for these CSPN if they were present, and late-M companions
would be detected for the closest of the sample.
The data do not constrain the presence of companions with lower mass,
such as from low-mass L or T dwarfs or planets.

\section{FUTURE IMPROVEMENTS}

Parallaxes from HST are likely to reach greater precision than the
USNO ground-based results.  However, limited HST observing time
means that an extensive HST program for planetary nebulae parallaxes
is unlikely to happen.  Therefore, dramatic improvements beyond the
ground-based program described here will have to wait until SIM or
GAIA are in use.

Several other PNe are likely to be at distances where useful parallax
measurements can be made now from the ground.  As well as repeating a
couple of the PNe in Table 3, it may be possible to get useful results
for NGC~246, NGC~1360, NGC~1514, NGC~3242, PG0108+101, A35, A36, LT5,
and Hu2-1.  Some of these objects may be added to our program soon.
Unfortunately, some of them have central stars (or companions
to the central stars) sufficiently bright
to require some change in the camera hardware in order to get higher
dynamic range to reach faint reference stars.  We hope other researchers
will include some of these objects in their studies to compare distances
determined with other methods.  If other PNe are of particular importance
and are likely to be at a distance closer than 500 pc, please contact us
to request their addition to the program.

\acknowledgments
We thank the referee, Dr. H. Smith, for helpful comments and for
suggesting adding Figure 4, as part of a thorough review.
This research has made use of the SIMBAD database,
operated at CDS, Strasbourg, France.

\clearpage

\begin{deluxetable}{lrrrcccccc}
% \rotate
\tabletypesize{\scriptsize}
\tablenum{1}
% \tablewidth{9.1 in}
\tablewidth{0pt}
\tablecaption{USNO Astrometric Data}
\tablehead{
\multicolumn{2}{c}{PN} &
\multicolumn{2}{c}{Number of} &
\colhead{Epoch} &
\colhead{$\pi_{rel}$} &
\colhead{$\Delta \pi$} &
\colhead{$\pi_{abs}$} &
\colhead{$\mu_{rel}$} &
\colhead{PA} \\
\colhead{Name} &
\colhead{PN G} &
\colhead{Frames} &
\colhead{Nights} &
\colhead{} &
\colhead{(mas)} &
\colhead{(mas)} &
\colhead{(mas)} &
\colhead{(mas yr$^{-1}$)} &
\colhead{(degrees)}
}
\startdata
\multicolumn{3}{l}{{\phn\phn}TI800 CCD Program:} &&&&&& \\
NGC 6853 & 060.8$-$03.6 & 146 & {\phn}68 & 88.4--95.6 &
  2.10 $\pm$ 0.38 & 0.53 $\pm$ 0.20 & 2.63 $\pm$ 0.43 &
  14.2 $\pm$ 0.3 &{\phn}60.5 $\pm$ 2.5 \\
NGC 6720 & 063.1+13.9   & 230 & 117 & 88.1--95.2 &
  0.58 $\pm$ 0.45 & 0.84 $\pm$ 0.32 & 1.42 $\pm$ 0.55 &
{\phn}9.0 $\pm$ 0.6 & 318.0 $\pm$ 3.3 \\
A74      & 072.7$-$17.1 & {\phn}53 & {\phn}50 & 88.5--94.8 &
  0.96 $\pm$ 0.62 & 0.37 $\pm$ 0.12 & 1.33 $\pm$ 0.63 &
{\phn}1.8 $\pm$ 0.3 & 296.6 $\pm$ 7.6 \\
Sh2-216  & 158.5+00.7   & 103 & {\phn}40 & 89.7--94.9 &
  6.31 $\pm$ 0.88 & 1.06 $\pm$ 0.47 & 7.37 $\pm$ 1.00 &
  23.8 $\pm$ 0.6 & 119.2 $\pm$ 1.5 \\
PuWe1    & 158.9+17.8   & {\phn}61 & {\phn}31 & 87.9--95.8 &
  2.58 $\pm$ 0.51 & 0.54 $\pm$ 0.24 & 3.12 $\pm$ 0.56 &
  21.0 $\pm$ 0.3 & 186.9 $\pm$ 0.5 \\
A21      & 205.1+14.2   & {\phn}66 & {\phn}53 & 87.9--95.8 &
  1.60 $\pm$ 0.50 & 0.25 $\pm$ 0.10 & 1.85 $\pm$ 0.51 &
{\phn}8.8 $\pm$ 0.5 & 204.0 $\pm$ 2.6 \\
\multicolumn{3}{l}{{\phn\phn}TEK2048 CCD Program:} &&&&&& \\
NGC 7293 & 036.1$-$57.1 & 228 &     169 & 92.6--02.6 &
  3.64 $\pm$ 0.47 & 0.92 $\pm$ 0.12 & 4.56 $\pm$ 0.49 &
  33.0 $\pm$ 0.1 &{\phn}86.7 $\pm$ 0.3 \\
NGC 6853 & 060.8$-$03.6 & 264 &     178 & 97.4--02.7 &
  3.21 $\pm$ 0.42 & 0.60 $\pm$ 0.20 & 3.81 $\pm$ 0.47 &
  13.0 $\pm$ 0.2 &{\phn}61.0 $\pm$ 1.0 \\
RE1738+665 & 096.9+32.0 & 204 &     137 & 97.3--03.5 &
  5.16 $\pm$ 0.40 & 0.75 $\pm$ 0.12 & 5.91 $\pm$ 0.42 &
  23.6 $\pm$ 0.1 & 130.4 $\pm$ 0.3 \\
DeHt5    & 111.0+11.6   & 187 &     162 & 96.8--05.9 &
  2.82 $\pm$ 0.54 & 0.52 $\pm$ 0.12 & 3.34 $\pm$ 0.56 &
  21.4 $\pm$ 0.2 & 214.6 $\pm$ 0.5 \\
PHL932   &125.9$-$47.0 &  57 &{\phn}40 & 03.7--06.0 &
  2.39 $\pm$ 0.61 & 0.97 $\pm$ 0.15 & 3.36 $\pm$ 0.62 &
  37.8 $\pm$ 0.4 &{\phn}74.2 $\pm$ 0.8 \\
HDW4     & 156.3+12.5   & 183 &     115 & 96.8--03.0 &
  3.86 $\pm$ 0.35 & 0.92 $\pm$ 0.20 & 4.78 $\pm$ 0.40 &
  24.4 $\pm$ 0.2 & 136.3 $\pm$ 0.4 \\
Sh2-216  & 158.5+00.7   & 280 &     163 & 92.8--01.0 &
  7.10 $\pm$ 0.24 & 0.71 $\pm$ 0.25 & 7.81 $\pm$ 0.35 &
  23.5 $\pm$ 0.1 & 119.6 $\pm$ 0.2 \\
PuWe1    & 158.9+17.8   & 249 &     122 & 92.8--01.0 &
  1.97 $\pm$ 0.34 & 0.60 $\pm$ 0.15 & 2.57 $\pm$ 0.37 &
  19.0 $\pm$ 0.2 & 184.2 $\pm$ 0.3 \\
Ton320   & 191.4+33.1   & 153 &     136 & 96.9--06.2 &
  1.23 $\pm$ 0.29 & 0.65 $\pm$ 0.15 & 1.88 $\pm$ 0.33 &
  11.7 $\pm$ 0.1 & 203.5 $\pm$ 0.3 \\
A7       & 215.5$-$30.8 & 172 &     155 & 93.9--03.0 &
  0.88 $\pm$ 0.37 & 0.60 $\pm$ 0.23 & 1.48 $\pm$ 0.42 &
  10.9 $\pm$ 0.1 & 304.5 $\pm$ 0.7 \\
A24      & 217.1+14.7   &  90 &{\phn}86 & 96.8--05.9 &
  1.45 $\pm$ 0.32 & 0.47 $\pm$ 0.10 & 1.92 $\pm$ 0.34 &
{\phn}3.5 $\pm$ 0.1 & 265.2 $\pm$ 2.6 \\
A31      & 219.1+31.2   & 135 &{\phn}95 & 96.9--03.0 &
  0.83 $\pm$ 0.26 & 0.92 $\pm$ 0.20 & 1.76 $\pm$ 0.33 &
  10.4 $\pm$ 0.1 & 226.5 $\pm$ 0.6 \\
PG1034+001 &        &  24 &{\phn}24 & 04.0--06.2 &
  3.68 $\pm$ 0.51 & 1.07 $\pm$ 0.15 & 4.75 $\pm$ 0.53 &
  87.9 $\pm$ 0.5 & 293.3 $\pm$ 0.5 \\
\enddata
% \tablenotetext{1}{Observations continuing.}
\end{deluxetable}

\clearpage

\begin{deluxetable}{lcccr}
\tablenum{2}
\tablewidth{0pt}
% \tablewidth{3.6 in}
\tablecaption{Photometry of the Central Stars$^1$}
\tablehead{
\colhead{Name} &
\colhead{$V$} &
\colhead{$B-V$} &
\colhead{$V-I$} &
\colhead{N}
}
\startdata
NGC 7293  &13.525 &$-$0.322 &$-$0.415 & 2 \\
NGC 6853  &13.989 &$-$0.298 &$-$0.392 & 3 \\
NGC 6720  &15.749 &$-$0.383 &$-$0.299 & 2 \\
A74       &17.046 &$-$0.334 &$-$0.259 & 2 \\
RE1738+665&14.580 &$-$0.342 &$-$0.339 & 3 \\
DeHt5     &15.474 &$-$0.221 &$-$0.166 & 2 \\
PHL932    &12.107 &$-$0.254 &$-$0.271 & 2 \\
HDW4      &16.528 &$-$0.221 &$-$0.236 & 1 \\
Sh2-216   &12.630 &$-$0.295 &$-$0.362 & 4 \\
PuWe1     &15.534 &$-$0.244 &$-$0.270 & 4 \\
Ton320    &15.702 &$-$0.277 &$-$0.408 & 2 \\
A21       &15.962 &$-$0.293 &$-$0.352 & 3 \\
A7        &15.479 &$-$0.283 &$-$0.330 & 2 \\
A24       &17.407 &$-$0.285 &$-$0.371 & 2 \\
A31       &15.519 &$-$0.285 &$-$0.314 & 2 \\
PG1034+001&13.211 &$-$0.299 &$-$0.418 & 1 \\
\enddata
\tablenotetext{1}{Johnson $B$ and $V$, and Cousins $I$ magnitudes.}
\end{deluxetable}

\clearpage

\begin{deluxetable}{lrcccl}
\tablenum{3}
\tablewidth{0pt}
% \tablewidth{7.0 in}
\tablecaption{Results}
\tablehead{
\colhead{PN} &
\colhead{$\pi_{abs}$} &
\colhead{Distance (1$\sigma$ Range)} &
\colhead{E($B-V$)} &
\colhead{M$_V$ (1$\sigma$ Range)} &
\colhead{V$_{\rm tan}$} \\
\colhead{} &
\colhead{(mas)} &
\colhead{(pc)} &
\colhead{} &
\colhead{} &
\colhead{(km s$^{-1}$)} 
}
\startdata
NGC 7293 & 4.56 $\pm$ 0.49 & 219 (198--246)& 0.03
  & 6.73 (6.47--6.94) & 34.2 $\pm$ 3.6 \\
NGC 6853$^1$& 2.64 $\pm$ 0.33 & 379 (337--433)& 0.07
  & 5.88 (5.59--6.14) & 24.1 $\pm$ 3.2 \\
NGC 6720 & 1.42 $\pm$ 0.55 & 704 (508--1149)& 0.08
  & 6.26 (5.20--6.97) & 30.0 $\pm$13.7 \\
A74      & 1.33 $\pm$ 0.63 & 752 (510--1428)& 0.12
  & 7.29 (5.90--8.14) &{\phn}6.4 $\pm$ 4.0 \\
RE1738+665& 5.91 $\pm$ 0.42 & 169 (158--182)& 0.05
  & 8.29 (8.12--8.43) & 18.9 $\pm$ 1.4 \\
DeHt5$^2$& 3.34 $\pm$ 0.56 & 300 (256--360)& 0.18
  & 7.53 (7.13--7.87) & 30.4 $\pm$ 5.3 \\
PHL932$^2$& 3.36 $\pm$ 0.62 & 298 (251--365)& 0.10
  & 4.43 (3.99--4.80) & 53.4 $\pm$10.2 \\
HDW4     & 4.78 $\pm$ 0.40 & 209 (193--228)& 0.16
  & 9.43 (9.24--9.61) & 24.2 $\pm$ 2.1 \\
Sh2-216  & 7.76 $\pm$ 0.33 & 129 (124--135)& 0.08
  & 6.83 (6.73--6.92) & 14.4 $\pm$ 0.6 \\
PuWe1    & 2.74 $\pm$ 0.31 & 365 (328--412)& 0.14
  & 7.28 (7.02--7.52) & 32.9 $\pm$ 3.8 \\
Ton320   & 1.88 $\pm$ 0.33 & 532 (452--645)& 0.07
  & 6.86 (6.44--7.21) & 29.5 $\pm$ 5.4 \\
A21      & 1.85 $\pm$ 0.51 & 541 (424--746)& 0.08
  & 7.05 (6.35--7.58) & 22.6 $\pm$ 6.7 \\
A7       & 1.48 $\pm$ 0.42 & 676 (526--943)& 0.10
  & 6.02 (5.30--6.56) & 34.9 $\pm$10.8 \\
A24      & 1.92 $\pm$ 0.34 & 521 (442--633)& 0.07
  & 8.61 (8.16--8.97) &{\phn}8.6 $\pm$ 1.6 \\
A31      & 1.76 $\pm$ 0.33 & 568 (478--699)& 0.07
  & 6.53 (6.08--6.91) & 28.0 $\pm$ 5.4 \\
PG1034+001$^2$& 4.75 $\pm$ 0.53 & 211 (189--237)& 0.05
  & 6.43 (6.18--6.67) & 87.9 $\pm$10.0 \\
\enddata
\tablenotetext{1}{Parallax is a weighted mean of the two USNO
measurements from Table 1 combined with the HST parallax from
Benedict et al. (2003) (see Sec. 2.3).}
\tablenotetext{2}{Results not final, observations continuing.}
\end{deluxetable}

\clearpage

\begin{deluxetable}{lcccccccc}
\tablenum{4}
\setlength{\tabcolsep}{0.03in}
\tablewidth{0pt}
% \tablewidth{6.8 in}
\tablecaption{Comparison with Other Studies}
\tablehead{
\colhead{PN} &
\multicolumn{3}{l}{Distance in pc:} &&&&& \\
\colhead{} &
\colhead{This} &
\colhead{{\phn\phn}CKS$^1${\phn\phn}} &
\colhead{TIW$^2$} &
\colhead{VZ$^3$} &
\colhead{Zhang} &
\colhead{Pottasch} &
\colhead{Spectroscopic$^4$} &
\colhead{Phillips} \\
\colhead{} &
\colhead{Paper} &
\colhead{1992} &
\colhead{1993} &
\colhead{1994} &
\colhead{1995} &
\colhead{1996} &
\colhead{2001} &
\colhead{2005a}
}
\startdata
NGC 7293  &219 &157 &{\phn}160&{\phn}400&{\phn}420 &280 &{\phn}290 &290 \\
NGC 6853  &379 &262 &{\phn}270&{\phn}400&{\phn}480 &360 &{\phn}440 &... \\
NGC 6720  &704 &872 &{\phn}840&       1000&1130&500 &1100      &... \\
A74       &752 &... &{\phn}230&...        &... &850 &1700      &600 \\
RE1738+665&169 &... & ...     &...        &... &200 &{\phn}200 &... \\
DeHt5     &300 &... &{\phn}400&...        &... &350 &{\phn}510 &510 \\
PHL932    &298 &819 &{\phn}590&       2340&3330&520 &{\phn}240 &520 \\
HDW4      &209 &... & ...     &...        &... &350 &{\phn}250 &250 \\
Sh2-216 &129&...&{\phn}{\phn}40&...       &... &130 &{\phn}190 &125 \\
PuWe1     &365 &141 &{\phn}240&...  &{\phn}900 &300 &{\phn}700 &433 \\
Ton320    &532 &... & $\ge$500&...        &... &350 &{\phn}350 &... \\
A21       &541 &... &{\phn}270&...        &... &500 &{\phn}630 &541 \\
A7        &676 &216 &{\phn}220&...  &{\phn}700 &550 &{\phn}700 &700 \\
A24       &521 &525 & $\ge$500&...        &1900&600 &...       &286 \\
A31       &568 &233 &{\phn}240&...        &1010&400 &1000      &326 \\
PG1034+001&211 &... & ...     &...        &... &... &{\phn}155 &... \\
\enddata
\tablenotetext{1}{Cahn {\it et al}. (1992)}
\tablenotetext{2}{Terzian (1993), based on Ishida \& Weinberger (1987)}
\tablenotetext{3}{Van de Steene \& Zijlstra (1994)}
\tablenotetext{4}{Napiwotzki (2001); Napiwotzki (1999);
                  Pottasch (1996) Table 6}
\end{deluxetable}

\clearpage

\begin{figure}
\plotone{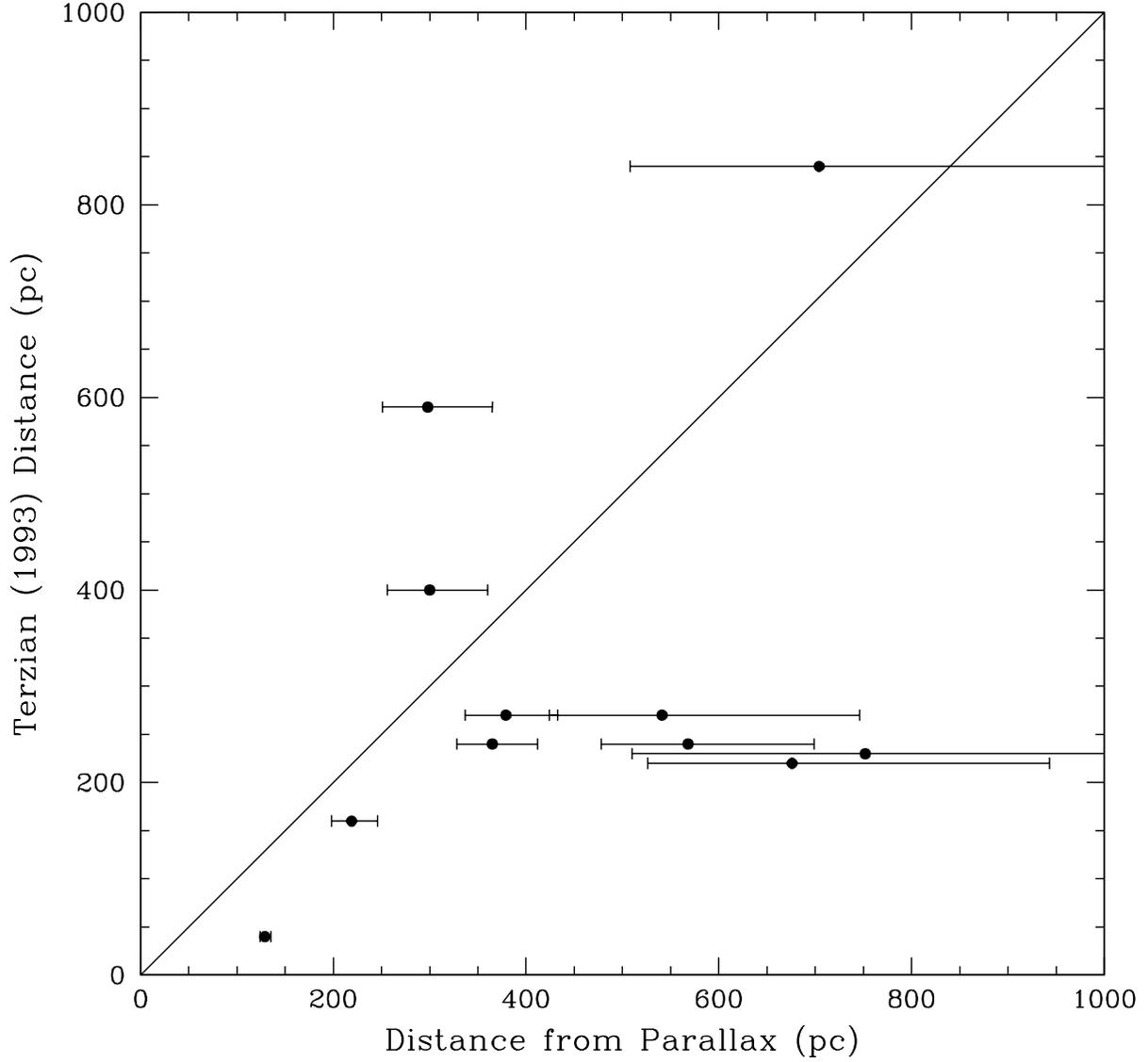}
\caption{Distances of nearby PNe tabulated by
Terzian (1993; his Table 3) and his primary source \citep{ish87}
compared with this paper.}
\end{figure}

\clearpage

\begin{figure}
\plotone{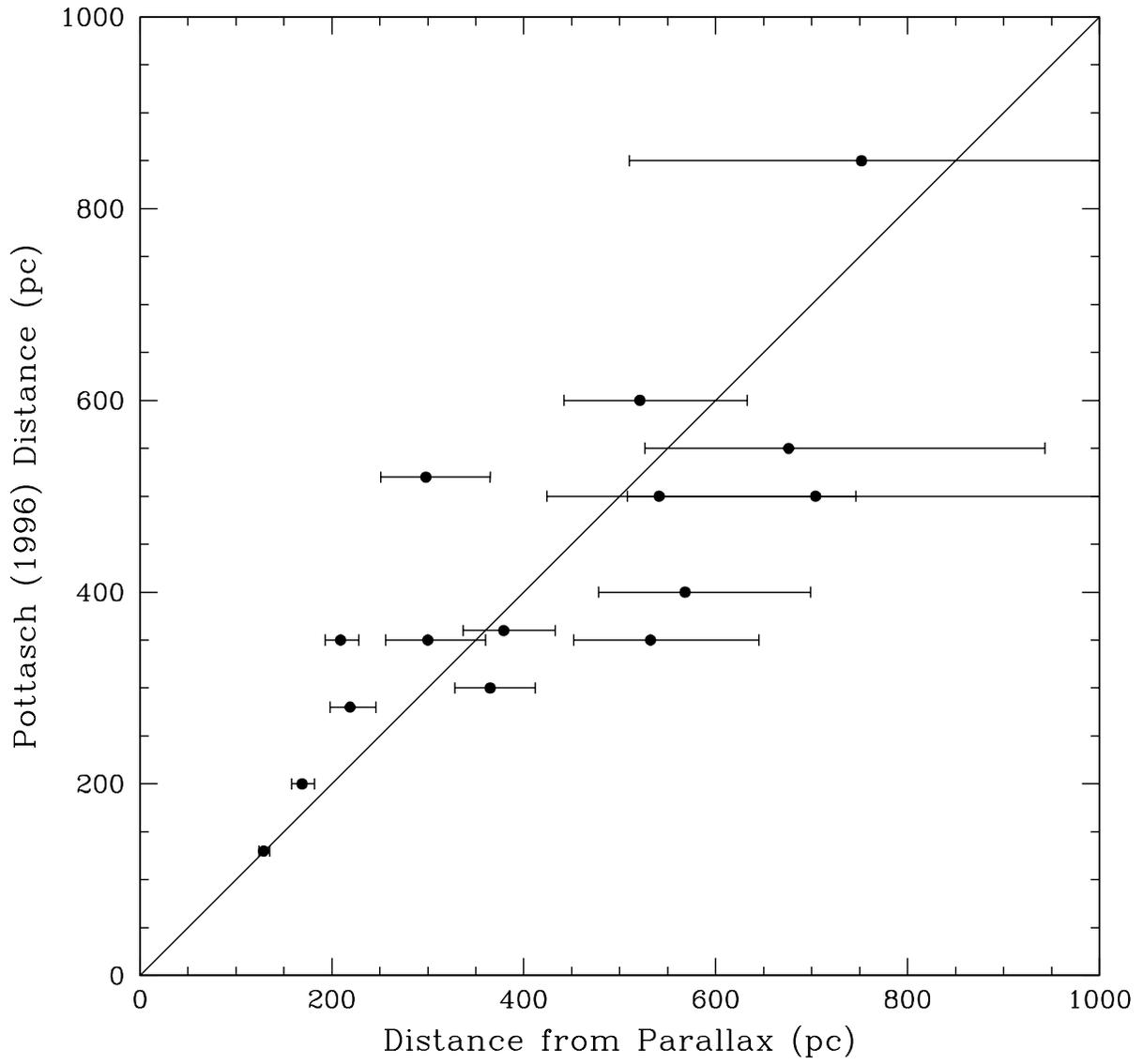}
\caption{Distances of PNe from Pottasch (1996;
his Table 9) compared with this paper.}
\end{figure}

\clearpage

\begin{figure}
\plotone{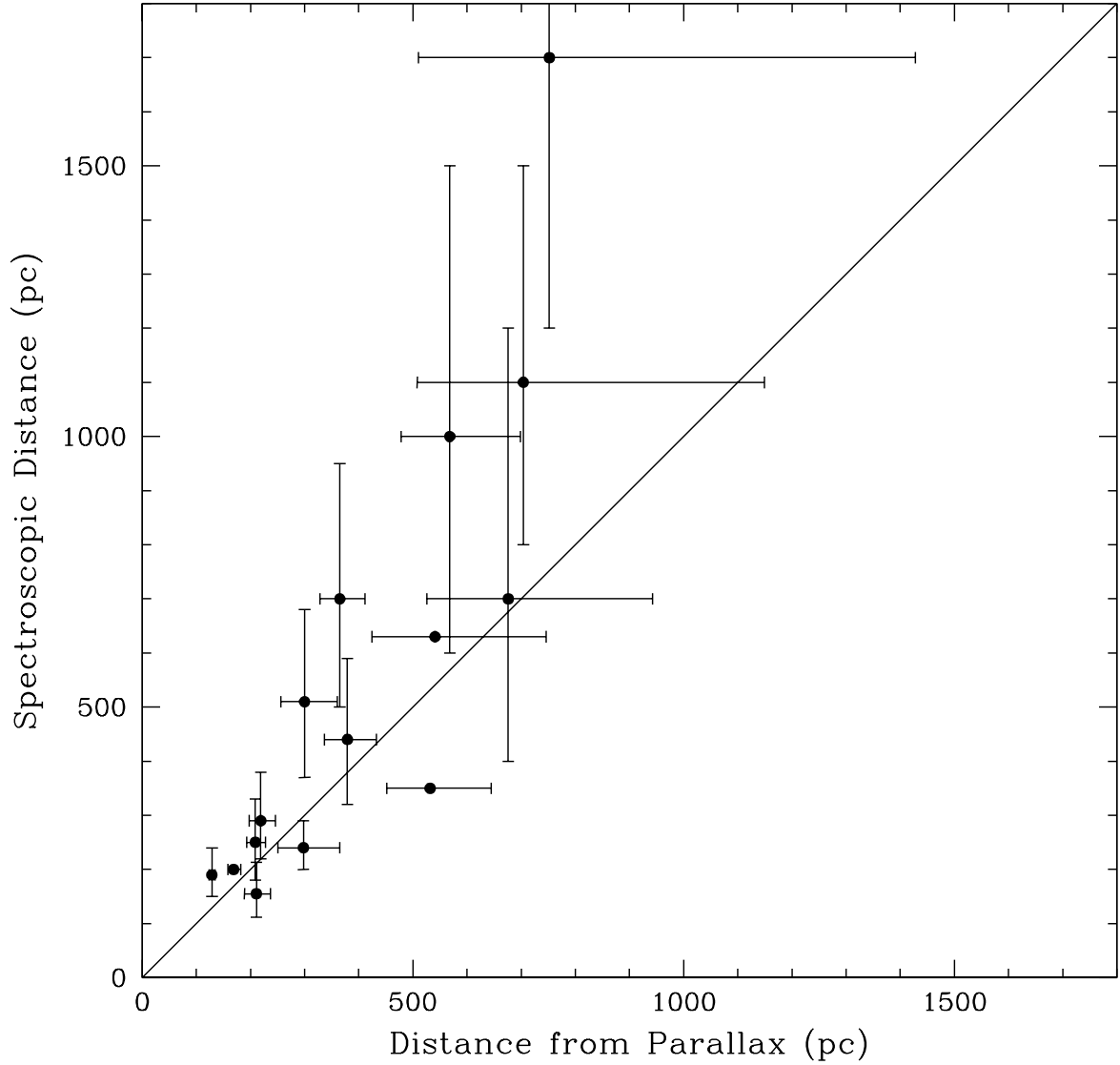}
\caption{Distances of PNe from spectroscopic analysis
of Napiwotzki (1999; 2001) plus two objects from Pottasch (1996;
his Table 6) with spectroscopic (``gravity'') distances
compared with this paper.}
\end{figure}

\clearpage

\begin{figure}
\plotone{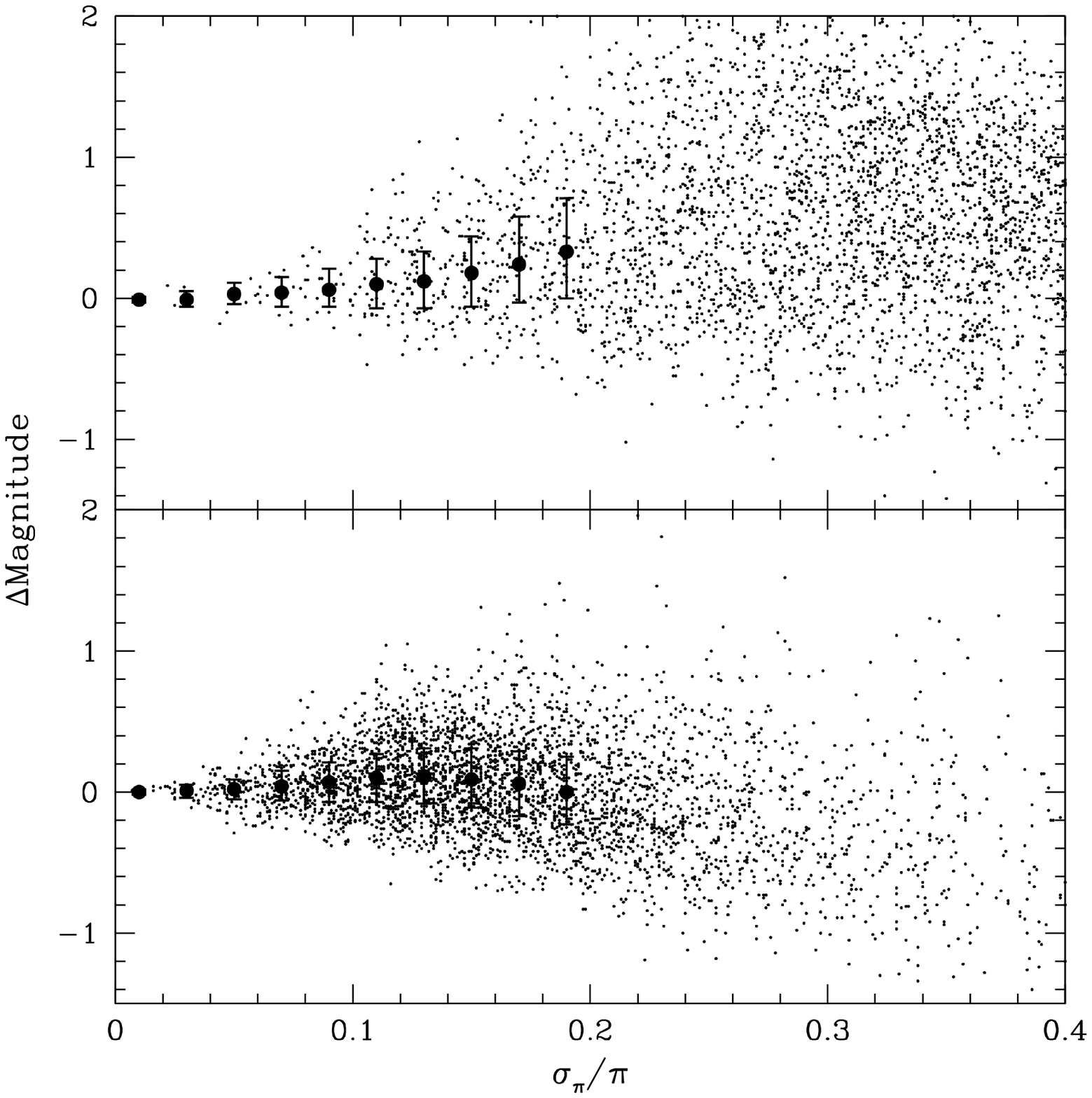}
\caption{The bias in distances determined from
parallaxes for a model disk population, expressed here as a bias
in absolute magnitudes.  The top panel shows the bias for a disk
population -- it is {\it not} representative of the sample
in this paper.  The lower panel includes a sample selection for
objects believed in advance to be nearby.  This panel is an attempt
to model the sample in this paper realistically.}
\end{figure}


\clearpage

\begin{thebibliography}{}

\bibitem[Acker et al.\ 1998]{ack98} Acker, A., Fresneau, A.,
   Pottasch, S.R., \& Jasniewicz, G. 1998, \aap, 337, 253
\bibitem[Barstow et al.\ 1994]{bar94} Barstow, M.A., Holberg, J.B.,
   Marsh, M.C., Tweedy, R.W., Burleigh, M.R., Fleming, T.A.,
   Koester, D., Penny, A.J., \& Sansom, A.E. 1994, \mnras, 271, 175
\bibitem[Barstow et al.\ 2003]{bar03} Barstow, M.A., Good, S.A.,
   Holberg, J.B., Hubeny, I., Bannister, N.P., Bruhweiler, F.C.,
   Burleigh, M.R., \& Napiwotzki, R. 2003, \mnras, 341, 870
\bibitem[Benedict et al.\ 2003]{ben03} Benedict, G.F., et al.
   2003, \aj, 126, 2549
\bibitem[Cahn et al.\ 1992]{cah92} Cahn, J.H., Kaler, J.B., \&
   Stanghellini, L. 1992, \aaps, 94, 399
\bibitem[Chu et al.\ 2004]{chu04} Chu, Y.-H., Gruendl, R.A.,
   Williams, R.M., Gull, T.R., \& Werner, K. 2004, \aj, 128, 2357
\bibitem[Ciardullo et al.\ 1999]{cia99} Ciardullo, R., Bond, H.E.,
   Sipior, M.S., Fullton, L.K., Zhang, C.Y., \& Schaefer, K.G.
   1999, \aj, 118, 488
\bibitem[Dahn et al.\ 2002]{dah02} Dahn, C.C., et al.\ 2002,
   \aj, 124, 1170
\bibitem[Dahn et al.\ 1973]{dah73} Dahn, C.C., Behall, A.L., \&
   Christy, J.W. 1973, \pasp, 85, 224
\bibitem[De Marco et al.\ 2004]{dem04} De Marco, O., Bond, H.E.,
   Harmer, D., \& Fleming, A.J.  2004, \apj, 602, L93
\bibitem[Frew \& Parker 2006]{fre06} Frew, D.J., \& Parker, Q.
   2006, in Proc. IAU Symp. 234, Planetary Nebulae in our Galaxy
   and Beyond, ed. M.J. Barlow \& R.H. M\'endez (Cambridge:
   Cambridge University Press), in press
\bibitem[Guti\'errez-Moreno et al.\ 1999]{gut99}
   Guti\'errez-Moreno, A., Anguita, C., Loyola, P., \& Moreno, H.
   1999, \pasp, 111, 1163
\bibitem[Hajian 2006]{haj06} Hajian, A.R. 2006, in Proc. IAU Symp.
   234, Planetary Nebulae in our Galaxy and Beyond, ed. M.J.
   Barlow \& R.H. M\'endez (Cambridge: Cambridge University Press),
   in press
\bibitem[Harrington \& Dahn 1980]{har80} Harrington, R.S., \&
   Dahn, C.C. 1980, \aj, 85, 454
\bibitem[Harris et al.\ 1997]{har97} Harris, H.C., Dahn, C.C.,
   Monet, D.G., \& Pier, J.R. 1997, in Proc. IAU Symp. 180,
   {\it Planetary Nebulae}, ed. H.J. Habing \& H.J.G.L.M Lamers
   (Dordrecht: Kluwer), 40
\bibitem[Harris et al.\ 2005]{har05} Harris, H.C., et al.\ 2005,
   in {\it Astrometry in the Age of the Next Generation of Large
   Telescopes}, ASP Conf. Ser. 338, ed. P.K. Seidelmann \&
   A.K.B. Monet (San Francisco: ASP), 122
\bibitem[Hewett et al.\ 2003]{hew03} Hewett, P.C., Irwin, M.J.,
   Skillman, E.D., Foltz, C.B., Willis, J.P., Warren, S.J., \&
   Walton, N.A. 2003, \apj, 599, L37
\bibitem[Ishida \& Weinberger 1987]{ish87} Ishida, K., \&
   Weinberger, R. 1987, \aap, 178, 227
\bibitem[Landolt 1992]{lan92} Landolt, A.U. 1992, \aj, 104, 340
\bibitem[Lisker et al.\ 2005]{lis05} Lisker, T., Heber, U.,
  Napiwotzki, R., Christlieb, N., Han, Z., Homeier, D., \&
  Reimers, D.  2005, \aap, 430, 223
\bibitem[Lutz \& Kelker 1973]{lut73} Lutz, T.E., \& Kelker, D.H.
   1973, \pasp, 85, 573
\bibitem[M\'endez et al.\ 1988]{men88} M\'endez, R.H., Groth,
   H.G., Husfeld, D., Kudritzki, R.-P., \& Herrero, A. 1988,
   \aap, 197, L25
\bibitem[Monet et al.\ 1992]{mon92} Monet, D.G., Dahn, C.C.,
   Vrba, F.J., Harris, H.C., Pier, J.R., Luginbuhl, C.B., \&
   Ables, H.D. 1992, \aj, 103, 638
\bibitem[Napiwotzki 1999]{nap99} Napiwotzki, R. 1999,
   \aap, 350, 101
\bibitem[Napiwotzki 2001]{nap01} Napiwotzki, R. 2001,
   \aap, 367, 973
\bibitem[Phillips 2005a]{phi05a} Phillips, J.P. 2005, \mnras, 357, 619
\bibitem[Phillips 2005b]{phi05b} Phillips, J.P. 2005, \mnras, 362, 847
\bibitem[Pier et al.\ 1993]{pie93} Pier, J.R., Harris, H.C.,
   Dahn, C.C., \& Monet, D.G. 1993, in Proc. IAU Symp. 155,
   {\it Planetary Nebulae}, ed. R. Weinberger \& A. Acker
   (Kluwer, Dordrecht), 175
\bibitem[Pottasch 1996]{pot96} Pottasch, S.R. 1996, \aap, 307, 561
\bibitem[Pottasch \& Acker 1998]{pot98} Pottasch, S.R. \&
   Acker, A. 1998, \aap, 329, L5
\bibitem[Rauch et al.\ 2004]{rau04} Rauch, T., Kerber, F., \&
   Pauli, E.-M. 2004, \aap, 417, 647
\bibitem[Smith 2003]{smi03} Smith, H., Jr. 2003, \mnras, 338, 891
\bibitem[Terzian 1993]{ter93} Terzian, Y. 1993, in Proc. IAU Symp.
   155, {\it Planetary Nebulae}, ed. R. Weinberger \& A. Acker
   (Kluwer, Dordrecht), 109
\bibitem[Tweedy 1995]{twe95} Tweedy, R.W. 1995, Nature, 373, 666
\bibitem[Tweedy \& Kwitter 1996]{twe96} Tweedy, R.W., \& Kwitter,
   K.B. 1996, \apjs, 107, 255
\bibitem[Van de Steen \& Zijlstra 1994]{van94}
   Van de Steene, G.C., \& Zijlstra, A.A. 1994, \aaps, 108, 485
\bibitem[van Altena et al.\ 1995]{van95} van Altena, W.F., Lee, J.T.,
   \& Hoffleit, E.D. 1995, The General Catalogue of Trigonometric
   Stellar Parallaxes, Fourth Edition (New Haven: Yale University
   Observatory)
\bibitem[Werner et al.\ 1995]{wer95} Werner, K., Dreizler, S., \&
   Wolff, B. 1995, \aap, 298, 567

\end{thebibliography}
\end{document}